# Group Delay Measurements of Multicore Fibers with Correlation Optical Time Domain Reflectometry

**Florian Azendorf**[1,2], *Member, OSA*, **Annika Dochhan**[1], *Member, IEEE*, **Florian Spinty**[1], **Mirko Lawin**[1], **Bernhard Schmauss**[2], **Michael Eiselt**[1], *Fellow, OSA, Senior Member, IEEE*

[1] *ADVA Optical Networking SE, Märzenquelle 1-3, 98617 Meiningen*
[2] *LHFT, Friedrich-Alexander-Universität Erlangen-Nürnberg, 91058 Erlangen, Germany*
E-Mail: Fazendorf@adva.com

**ABSTRACT**
Several multi-core fibers (MCF) were characterized using Correlation Optical Time Domain Reflectometry (C-OTDR) in terms of propagation delay and polarization mode dispersion (PMD). The results show that the propagation delay in the cores depends on the position of the core in the fiber and that the differential delay between the cores varies with temperature.

## 1. INTRODUCTION

The group delay of an optical signal in the fiber becomes a critical parameter for future 5G networks. For some applications, it is not the absolute delay value that is of concern, but rather the differential delay between two optical paths. Synchronization applications require a stable and symmetric group delay between the master clock and slave clock and vice versa. The knowledge of the differential delay between paths is important. In current synchronization applications, like IEEE 1588v2 (PTP), an asymmetry of a few nanoseconds between the master and slave clock is sufficient. To achieve a better synchronization in future applications, an asymmetry reduced by an order of magnitude will be required. Another application is the transmission of radio over fiber signals and the use of phase array antennas [1,2]. The allowed phase error of 30 degrees at 26 GHz leads to a maximum skew of 3.2 ps between the paths. Environmental temperature changes are the main cause of changes in propagation delay. A typical unjacketed standard single mode fiber has a temperature delay coefficient (TDC) of 7.49 ppm/K [3]. While four fibers in the same conduit will experience approximately the same environmental temperature, delay differences on the order of up to 800 ps have been observed in 8 km of jacketed fiber [4]. In multi-core fiber, the group delay of all cores is expected to be the same. Therefore, to reduce the differences between two optical paths, multi-core fiber can be considered as advantageous compared to using multiple separate single mode fibers. Due to the fiber design, however, propagation delay differences, or skew, between the fiber cores have also been observed [5,6]. While fixed delay differences could be characterized during fiber installation and be compensated for skew sensitive applications, temperature changes might lead to additional skew variations. To obtain an insight into the temperature dependent skew of multi-core fiber, the cores must be characterized simultaneously to avoid temperature fluctuations during consecutive measurements. In this paper, Correlation-Optical Time Domain Reflectometry (C-OTDR) was used to characterize the group delay of four different multi-core fiber samples over temperature. In addition, using the modulation phase shift (MPS) method [7], all fibers were characterized with respect to polarization mode dispersion. While the MPS method requires access to both ends of the fiber under test (FuT) and each core was measured consecutively, the C-OTDR measures the group delay of an optical fiber from one end with a precision in the order of picoseconds [8]. With this method, four cores can be measured simultaneously to eliminate the impact of temperature fluctuations on the skew measurement. The measurement technique was integrated in a portable unit (30 cm x 30 cm), based on a multi-processor-system-on-chip integrated in a field-programmable-gate-array (MPSoC-FPGA), using a small formfactor pluggable (SFP) 2.5G transceiver for the optical front end.

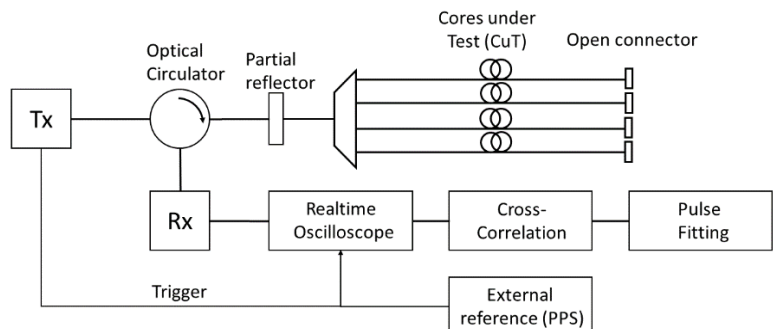

Fig. 1 Schematic of the C-OTDR laboratory setup to simultaneously measure four cores of a multi-core fiber.

## 2. C-OTDR METHOD USING LABORATORY SETUP

The typical OTDR is limited in time resolution due to the trade-off between pulse width and signal amplitude. One solution to increase the time resolution is to send bursts of bit sequences instead of pulses. and correlating the backscattered or reflected signal with the transmitted bit sequence. This method is called C-OTDR. Fig. 1 shows the schematic of the C-OTDR. Via an optical circulator, a 10-Gbit/s bit sequence burst was sent into the fiber. The burst followed by a fill pattern was generated with an arbitrary waveform generator and modulated with a Mach-Zehnder modulator on a continuous wave laser signal at 1550 nm. The optical signal was reflected by a partial reflecting connector to obtain a time reference. Afterwards, the signal was split between four cores that are simultaneously characterized. At the distal end of each core, an open connector or a highly reflective device was used to create a defined reflection from the core end. The reflected signals were received with a PIN/TIA combination and recorded by a real-time oscilloscope, sampling with 50 GS/s and averaging over e.g. 1000 traces. The real-time oscilloscope was synchronized with the burst trigger of the burst source and referenced to a precise time base. The recorded signal was then processed offline. As a first step, the signal was correlated with the transmitted bit sequence, resulting in correlation peaks with a width of approximately 80 ps and a sample resolution of 20 ps. Fig. 2a shows the cross-correlation peaks from the end reflections of four cores after approximately 5 km of fiber. The reflection peaks were then fitted with a Gaussian function, whose peak position was taken as the reflection point, resulting in a time resolution of better than one sample period, as shown in Fig. 2b.

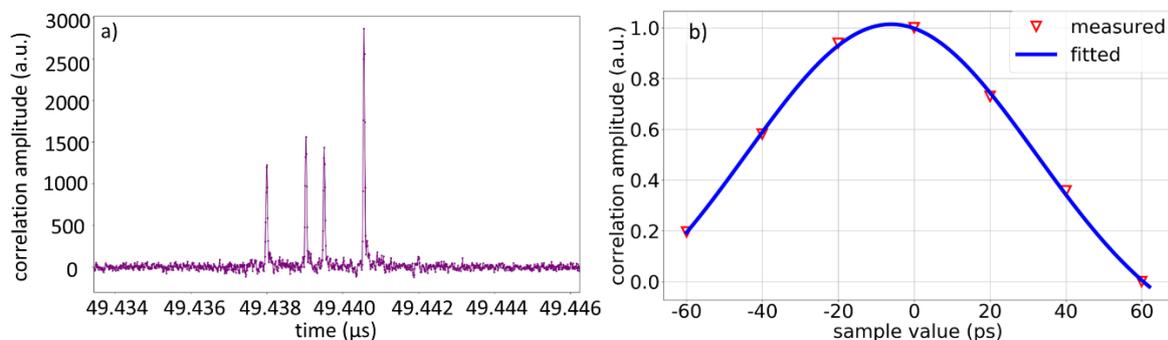

Fig. 2 Measurement results of 4-core measurement: a) Correlation of end reflection signals of four cores of 5 km fiber, b) fit of Gaussian shape to one correlation peak.

## 3. MULTI-CORE FIBER CHARACTERIZATION

### 3.1 Group delay measurements

The group delay of all cores of two 7-core fibers and two 19-core fibers from different vendors was characterized using the C-OTDR method. Four cores, including the center core, were simultaneously characterized and the results were referenced to the delay measured for the center core. This enabled a compensation for environmental fluctuations during the measurement. For comparison, the group delay was also characterized using the modulation phase shift (MPS) method, which only allowed the characterization of one core at a time. As only consecutive measurements were possible, delay variations of up to 300 ps were observed in comparison to the C-OTDR technique. The differential delay, or skew, of all cores was calculated with respect to the center core. In Fig. 3, as an example, the cross section of a 5-km, 19-core fiber is shown with the labels showing the skew of each core with respect to the center core #10.

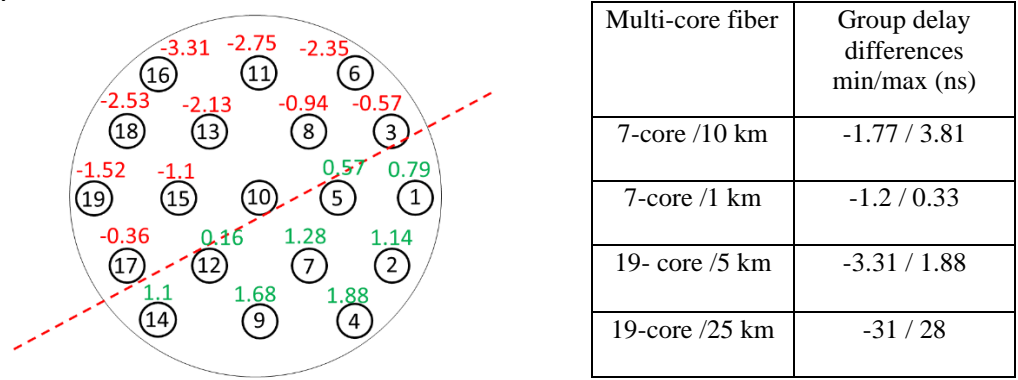

| Multi-core fiber | Group delay differences min/max (ns) |
|---|---|
| 7-core /10 km | -1.77 / 3.81 |
| 7-core /1 km | -1.2 / 0.33 |
| 19- core /5 km | -3.31 / 1.88 |
| 19-core /25 km | -31 / 28 |

Fig. 3 Left: Differential group delay variations with respect to the center core of the 5-km, 19-core fiber; right: minimum and maximum group delay differences of different multi-core fibers.

It can be seen that the skew between the cores depends on the position of the core in the fiber. One side of the fiber cross area experienced a higher delay than the opposite area. We assume that this effect is caused by spooling of the fiber.

### 3.2 Temperature dependent group delay and skew over temperature

It was mentioned above that temperature has a major impact on the group delay. To obtain an insight how the temperature impacts also the skew between the cores, all fibers were placed into a temperature-controlled cabinet. A group of four cores was measured simultaneously over a temperature range of 10 °C to 50 °C in steps of 10 K. To allow the temperature to fully settle, we waited for 1.5 hours after each temperature step. Fig. 4a shows the evolution of the group delay over temperature of the 5-km, 19-core fiber. The behavior over temperature for each fiber was quite linear with a temperature delay coefficient (TDC) of 7.1 to 7.49 ppm/K. This behavior is comparable to an unjacketed single mode fiber. From the results of the temperature dependent group delay, the skew over temperature with respect to the central core was calculated. Over a temperature range of 40 K, the skew, when referenced to 10 degC, exhibited a peak-to-peak variation of up to 25 ps, as shown in Fig. 4b, but no consistent trend could be observed. Fig. 4c shows a similar trend of the skew over temperature with peak to peak variations of 60 ps for a 10-km, 7-core fiber. All fibers investigated in this paper showed no consistent trend over a temperature range of 40K with variations from -38 ps to 25 ps. These variations might be caused by PMD of the fiber, as temperature changes would impact the random polarization evolution and could impact the group delay by several times the PMD value. Therefore, the PMD of the MCF was also investigated.

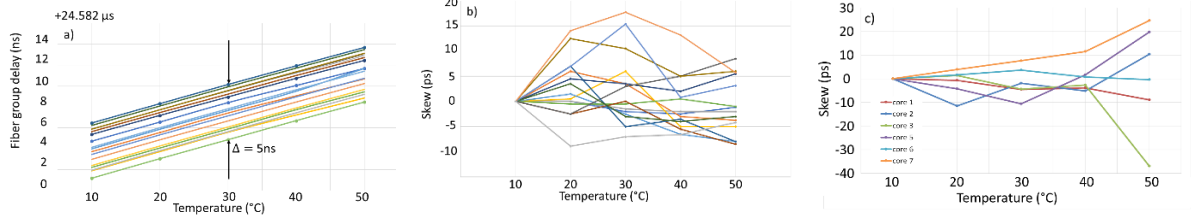

Fig. 4 a) Evolution of the fiber group delay of 5-km, 19-core MCF over a temperature range of 40 K; temperature dependent skew of b) 19-core and c) 7-core fiber with respect to the central core, normalized to the skew at 10°C.

### 3.3 PMD measurements

The differential group delay (DGD) of the fiber cores was measured, using the MPS method for four different input states of polarization, as a function of wavelength over a spectral bandwidth of 110 nm. The PMD was calculated as the average value of the DGD. The results showed high PMD values (up to 22 ps) for either the edge cores or all cores of the multi-core fibers. The central core of all multicore fibers showed PMD values below 0.43 ps. We assume that these high PMD values are caused by birefringence, induced due to the non-symmetric stress from neighbor cores. This assumption can be supported by the observation that the cores of the inner ring of the 5-km, 19-core fiber, with neighbors allocated symmetrically, showed low PMD, whereas the edge cores exhibited PMD values up to 6 ps. Fig. 5 shows the core distribution of the 10-km, 7-core and the 5-km, 19-core fibers with the PMD values in ps for each core. In Fig. 5c, the average PMD values are shown for all fibers.

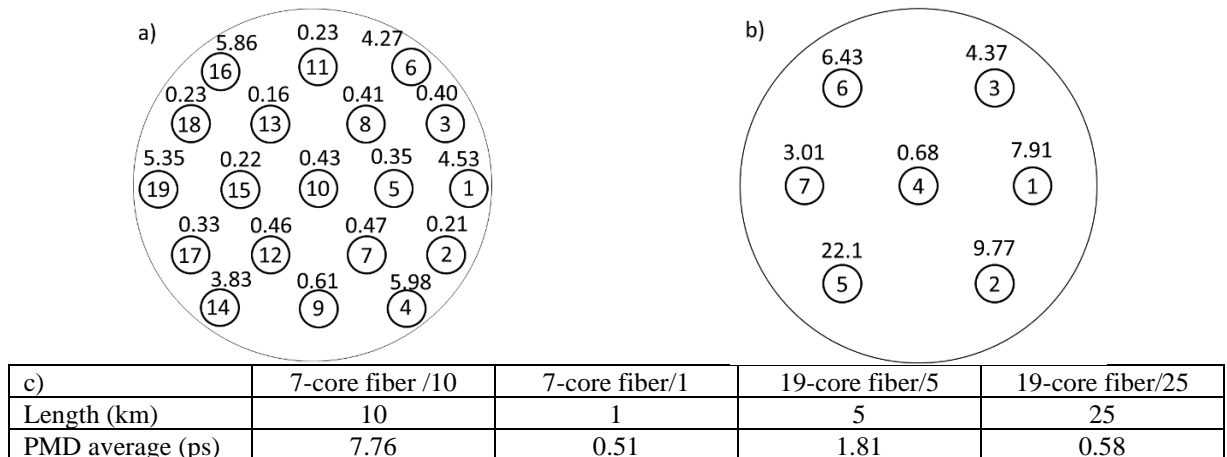

| c) | 7-core fiber /10 | 7-core fiber/1 | 19-core fiber/5 | 19-core fiber/25 |
|---|---|---|---|---|
| Length (km) | 10 | 1 | 5 | 25 |
| PMD average (ps) | 7.76 | 0.51 | 1.81 | 0.58 |

Fig. 5 Core distribution of a) 5-km, 19-core fiber and b) 10-km, 7-core fiber with measured PMD values in ps; c) average PMD of different multi-core fibers.

## 4. MPSOC-FPGA BASED SETUP

To enable mobile measurements, the C-OTDR setup described in section 2 was integrated using an SFP transceiver as optical front-end and a multi-processor-system-on-chip integrated in a field-programmable-gate-array (MPSoC-FPGA) for signal recording and processing. The receive signal chain consists of an avalanche photodiode (APD) with a transimpedance amplifier and a limiting amplifier. In this setup, the 7-bit analog to digital converter of the real-time oscilloscope was substituted by a 1-bit slicer, implemented as the combination of the limiting amplifier in the SFP and the high-speed data input to the FPGA. Averaging over several received signal traces results in an increased amplitude resolution, as discussed in the following.

Multiple reflection points in the fiber would result in a superposition of the reflected signals at the photodiode. Because the system used in these experiments only uses a one-bit analog to digital converter, it is not possible to distinguish between one or multiple reflected signals received simultaneously. To mitigate this problem and to improve the amplitude resolution, multiple received traces are accumulated. As noise is superimposed to the reflected signal, stemming from backscattering with random optical phases and from the thermal noise of the APD/TIA combination, a "high" bit decision is made by the one-bit slicer more often for a large reflected signal power than for a small reflected power. After adding up multiple received signal traces, all affected by noise, the sum of the "high" bits at a time slot is a measure of the reflected power at this time. Having increased the amplitude resolution in this way, all superimposed reflections can be distinguished in the following correlation. To achieve a fixed phase alignment between the sampling clock and the transmitted bit sequence burst for all accumulated bursts, a trigger bit is asserted on sending out the first bit of each bit sequence. Upon asserting this bit, the receiver starts sampling the incoming data from the SFP into memory, summing up the received bits for each time slot relative to the trigger. The amplitude of the sampled signal can then be measured by the number of accumulated "high" bits. After summing up to a predefined number of received traces, the sums for each time slot get transferred to the CPU in the MPSoC, where the data processing is performed.

## 5. CONCLUSIONS

Several multi-core fibers were characterized by using the correlation OTDR, showing group delay differences of 0.5 to 2 ns/km between the fiber cores. All fibers showed positive and negative skew values relative to the central core, depending on the position of the cores in the fiber cross section. We assume that these differences are caused by the spooling of the fiber. For skew sensitive applications, these delay differences could be compensated during the deployment. The skew variations over temperature showed variations of up to 20 ps, which might be due to PMD, as all fibers showed high PMD values for at least some cores. For skew sensitive applications, the skew over temperature needs a precise monitoring and can only be compensated, if the skew is not based on PMD. With a precise monitoring of the delay differences, multi-core fibers are therefore a candidate for future skew sensitive applications. We also reported in this paper on the integration of the correlation OTDR measurement technique into an MPSoC-FPGA with an SFP transceiver. Signal averaging improved the resolution of the one-bit slicer.

## ACKNOWLEDGEMENTS
This work was partially funded by the blueSPACE project with funding from the European Union's Horizon 2020 research and innovation programme under grant agreement number 762055 and by the German Federal Ministry of Education and Research (BMBF) under the project OptiCON with grant agreement number 16KIS0989K.

## REFERENCES
[1] Chris Roeloffzen, et al., "Low loss Si3N4 TriPleX optical waveguides: technology and application overview" IEEE JSTQE, 24(4), 2018.
[2] Thiago Raddo, et al. "Analog Radio-over-Fiber 5G Fronthaul Systems: blueSpace and 5G-PHOS Projects Convergence", EUCNC 2019, Valencia Spain, June 2019.
[3] G. Lutes and W. Diener, "Thermal coefficient of delay for various coaxial and -optic cables," TDA Progress Report 42-99, Nov. 1989, available online at https://ipnpr.jpl.nasa.gov/progress_report/42-99/99E.PDF
[4] F. Azendorf, et al., „Long-Term Latency Measurement of Deployed Fiber, OFC 2019, San Diego, March 2019.
[5] B. J. Puttnam, et al., „ Inter-Core Skew Measurements in Temperature Controlled Multi-Core Fiber" OFC 2018, San Diego, March 2018.
[6] F. Azendorf, et al., „Characterization of Multi-Core Fiber Group Delay with Correlation OTDR and Modulation Phase Shift Methods" OFC 2020, San Diego, March 2020.
[7] Agilent 86038B Photonic Dispersion and Loss Analyzer User's Guide available online at https://literature.cdn.keysight.com/litweb/pdf/86038-90B03.pdf?id=726642.
[8] F. Azendorf, et al., „Improvement of accuracy for measurement of 100-km fibre latency with Correlation OTDR", ECOC 2019, Dublin, September 2019.